\documentclass[12pt]{article}
\usepackage{graphicx}

\newsavebox{\sboxpubnumber}
\newsavebox{\sboxpubdate}
\newcommand{\pubdate}[1]{\begin{lrbox}{\sboxpubdate}{#1}\end{lrbox}}
\newcommand{\pubnumber}[1]{\begin{lrbox}{\sboxpubnumber}{\begin{tabular}{l} 
#1 \\
				 \usebox{\sboxpubdate}
				 \end{tabular}}
                           \end{lrbox}
                           \pubblock}
\newcommand{\Title}[1]{\begin{center} {\Large #1 } \end{center}}
\newcommand{\Author}[1]{\begin{center}{ \sc #1} \end{center}}
\newcommand{\Address}[1]{\begin{center}{ \it #1} \end{center}}

\newcommand{\pubblock}{\rightline{
			\usebox{\sboxpubnumber}}}
\newenvironment{Abstract}{\begin{quotation}  }{\end{quotation}}
\newenvironment{Presented}{\begin{quotation} \begin{center}
             PRESENTED AT\end{center}\bigskip
      \begin{center}\begin{large}}{\end{large}\end{center}
      \end{quotation}}

\def\gtwid{\mathrel{\raise.3ex\hbox{$>$\kern-1.05em\lower1ex\hbox{
$\sim$}}}}
\def\ltwid{\mathrel{\raise.3ex\hbox{$<$\kern-1.05em\lower1ex\hbox{
$\sim$}}}}

\begin{document}

\begin{titlepage}
\pubdate{\today}                    
\pubnumber{UFIFT-HEP-01-24} 

\vfill
\Title{The Big Flow}
\vfill
\Author{Pierre Sikivie\footnote{This work is supported in part by the 
U.S. Department of Energy under grant DE-FG02-97ER41029.}}
\Address{Department of Physics, University of Florida\\
         Gainesville, FL 32611, USA}
\vfill
\begin{Abstract}
The late infall of cold dark matter onto an isolated galaxy, such as our
own, produces streams and caustics in its halo.  The outer caustics are
topological spheres whereas the inner caustics are rings.  The
self-similar model of galactic halo formation predicts that the caustic
ring radii $a_n$ follow the approximate law $a_n \sim 1/n$.  In a study
of 32 extended and well-measured external galactic rotation curves
evidence was found for this law.  In the case of the Milky Way, the
locations of eight sharp rises in the rotation curve fit the prediction
of the self-similar model at the 3\% level.  Moreover, a triangular
feature in the IRAS map of the galactic plane is consistent with the
imprint of a ring caustic upon the baryonic matter.  These observations
imply that the dark matter in our neighborhood is dominated by a single
flow.  Estimates of that flow's density and velocity vector are given.
\end{Abstract}
\vfill
\begin{Presented}
    COSMO-01 \\
    Rovaniemi, Finland, \\
    August 29 -- September 4, 2001
\end{Presented}
\vfill
\end{titlepage}
\def\thefootnote{\fnsymbol{footnote}}
\setcounter{footnote}{0}

\section{Introduction}

There are compelling reasons to believe that the dark matter of the
universe is constituted in large part of non-baryonic collisionless
particles with very small primordial velocity dispersion, such as
axions and/or weakly interacting massive particles (WIMPs) \cite{CDM}.  
Generically, such particles are called cold dark matter (CDM).  Knowledge
of the distribution of CDM in galactic halos, and in our own halo in
particular, is of paramount importance to understanding galactic 
structure and predicting signals in experimental searches for dark 
matter.

One should expect this dark matter to form caustics.  A caustic is a 
place in physical space where the density is very large because the 
sheet on which the dark matter particles lie in phase-space has a 
fold there.  Caustics are commonplace in the propagation of light.  
An instructive example is given by the sharp luminous lines at the 
bottom of a swimming pool on a breezy sunny day.  Two conditions 
must be satisfied for caustics to occur generically.  First, the 
propagation must be collisionless.  Second the flow must have low 
velocity dispersion.  Light propagation is collisionless, and the 
flow of light from a point source has zero velocity dispersion.  Thus 
caustics are common in light.  Caustics in ordinary matter are very 
unusual because ordinary matter is not normally collisionless.  But 
CDM is collisionless and has very small velocity dispersion.  This 
guarantees that caustics are common in the distribution of CDM.

The primordial velocity dispersion of the cold dark matter candidates 
is indeed very small, of order 
\begin{equation}
\delta v_a (t) \sim 3\cdot 10^{-17} \left( {10^{-5} eV\over
m_a}\right)~\left({t_0\over t}\right)^{2/3}
\end{equation}
for axions, and 
\begin{equation}
\delta v_W (t) \sim 10^{-11} \left({GeV\over m_W}\right)^{1/2}
~\left({t_0\over t}\right)^{2/3}
\end{equation}
for WIMPs.  Here $t_0$ is the present age of the universe and 
$m_a$ and $m_W$ are respectively the masses of the axion and WIMP.   
The small velocity dispersion means that the dark matter particles 
lie on a thin 3-dim. sheet in 6-dim. phase-space.  The thickness of 
the sheet is $\delta v$.  The sheet cannot break and hence its 
evolution is constrained by topology.  

Where a galaxy forms, the sheet wraps up in phase-space,
turning clockwise in any two dimensional cut $(x, \dot{x})$ of that
space.  $x$ is the physical space coordinate in an arbitrary direction and
$\dot{x}$ its associated velocity.  The outcome of this process is a
discrete set of flows at any physical point in a galactic halo \cite{ips}.  
Two flows are associated with particles falling through the galaxy for 
the first time ($n=1$), two other flows are associated with particles
falling through the galaxy for the second time ($n=2$), and so
on.  Scattering in the gravitational wells of inhomogeneities in 
the galaxy (e.g. molecular clouds and globular clusters) are 
ineffective in thermalizing the flows with low values of $n$.  

Caustics appear wherever the projection of the phase-space sheet onto
physical space has a fold \cite{cr,lens,Tre,sing}.  Generically, caustics
are surfaces in physical space.  On one side of the caustic surface
there are two more flows than on the other.  At the surface, the dark
matter density is very large.  It diverges there in the limit of zero
velocity dispersion.  There are two types of caustics in the halos of
galaxies, inner and outer.  The outer caustics are topological spheres
surrounding the galaxy.  They are located near where a given outflow
reaches its furthest distance from the galactic center before falling
back in.  The inner caustics are rings \cite{cr}.  They are located
near where the particles with the most angular momentum in a given
inflow reach their distance of closest approach to the galactic center
before going back out.  A caustic ring is a closed tube whose
cross-section is a $D_{-4}$ (also called {\it elliptic umbilic})
catastrophe \cite{sing}.  The existence of these caustics and their
topological properties are independent of any assumptions of symmetry.

As was mentioned earlier, the primordial velocity dispersion of the
leading cold dark matter candidates is extremely small.  However,             
to a coarse-grained observer, the dark matter falling onto a galaxy 
may have additional velocity dispersion because the phase-space 
sheet on which the dark matter particles lie may be wrapped up 
on scales which are small compared to the galaxy as a whole.  This
effective velocity dispersion is associated with the clumpiness of 
the dark matter before it falls onto the galaxy.  For the caustics 
in a galaxy not to be washed out, the effective velocity dispersion 
of the infalling dark matter must be much less than the rotation 
velocity of the galaxy, say less than 30 km/s for our galaxy.  However, 
an upper bound of order 50 m/s can be obtained from observation, as 
explained below.

Primordial peculiar velocities are expected to be the same for baryonic
and dark matter particles because they are caused by gravitational forces.
Later the velocities of baryons and CDM differ because baryons collide
with each other whereas CDM is collisionless. However, because angular
momentum is conserved, the net angular momenta of the dark matter and
baryonic components of a galaxy are aligned.  Since the caustic rings
are located near where the particles with the most angular momentum in
a given infall are at their closest approach to the galactic center,
they lie close to the galactic plane.

\section{Caustic ring radii}

A specific proposal has been made for the radii $a_n$ of caustic rings
\cite{cr}:
\begin{equation}
\{a_n: n=1,2, ...\} \simeq (39,~19.5,~13,~10,~8,...) {\rm kpc}
\times \left({j_{\rm max}\over 0.25}\right) \left({0.7\over h}\right)
\left({v_{\rm rot} \over 220 {{\rm km} \over {\rm s}}} \right)
\label{crr}
\end{equation}
where $h$ is the present Hubble constant in units of
$100\,{\rm km/(s~Mpc)}$, $v_{\rm rot}$ is the rotation velocity of the
galaxy and $j_{\rm max}$ is a parameter with a specific value for each
halo.  For large $n$, $a_n \sim 1/n$.  Eq. \ref{crr} is predicted by
the self-similar infall model \cite{ss,sty} of galactic halo formation.
$j_{\rm max}$ is then the maximum of the dimensionless angular momentum
$j$-distribution \cite{sty}.  The self-similar model depends upon a
parameter $\epsilon$ \cite{ss}.  In CDM theories of large scale structure
formation, $\epsilon$ is expected to be in the range 0.2 to 0.35
\cite{sty}.
Eq. \ref{crr} is for $\epsilon = 0.3$.  However, in the range
$0.2 < \epsilon < 0.35$, the ratios $a_n/a_1$ are almost
independent of $\epsilon$.  When $j_{\rm max}$ values are quoted
below, $\epsilon = 0.3$ and $h = 0.7$ will be assumed.

Since the caustic rings lie close to the galactic plane, they cause
bumps in the rotation curve, at the locations of the rings.  In
ref. \cite{kinn} a set of 32 extended well-measured rotation curves
was analyzed and statistical evidence was found for bumps distributed
according to Eq. \ref{crr}.  That study suggests that the $j_{\rm max}$
distribution is peaked near 0.27.  The rotation curve of NGC3198, one
of the best measured, by itself shows three faint bumps which are
consistent with Eq. \ref{crr} and $j_{\rm max} = 0.28$.

A recent paper \cite{mw} gives evidence for ring caustics in our 
own galaxy.

\section{Ring caustics in the Milky Way}

A detailed north inner galactic rotation curve was obtained \cite{clem}
from the Massachusetts-Stony Brook Galactic Plane CO survey \cite{CO}. It
exhibits a series of eight sharp rises in the range of (galactocentric)
radii 3 to 7 kpc.  For each, Table I lists the radius $r_1$ where the
rise starts, the radius $r_2$ where it ends, and the increase $\Delta
v$ in rotation velocity.  The rises are interpreted here as due to the
presence of caustic rings of dark matter in the galactic plane.  Each
$r_1$ should therefore be identified with a caustic ring radius $a_n$,
and $r_2 - r_1$ with the caustic ring width $p_n$ \cite{sing}.  The
ring widths depend in a complicated way on the velocity distribution
of the infalling dark matter at last turnaround \cite{sing} and are
not predicted by the model.  They also need not be constant along the
ring.  In Table I, the numbers in parentheses are for two less distinct
rises between 7 kpc and our own radius $r_\odot$, taken to be 8.5 kpc.

The fourth column shows the caustic ring radii $a_n^{\rm I}$ of the
$\epsilon = 0.3$ self-similar infall model fitted to the eight
rises between 3 and 7 kpc, assuming that these are due to caustic
rings $n = 7 ... 14$ (fit I).  This is a one-parameter ($j_{\rm max}$)
fit minimizing $rmsd \equiv [{1 \over 8} {\displaystyle \sum_{n=7}^{14}}
( 1 - {a_n \over r_{1 n}})^2]^{1 \over 2}$.  It yields $j_{\rm max} =
0.263$ and $rmsd = 3.1 \%$.  The fifth column shows the radii
$a_n^{\rm II}$ assuming that the eight rises between 3 and 7 kpc
are due to caustic rings $n = 6 ... 13$ (fit II).  In this case
$j_{\rm max} = 0.239$ and $rmsd = 2.8 \%$.  Fits of similar quality
are obtained for the other values of $\epsilon$ in the range 0.20 to
0.35, or by assuming simply $a_n \sim 1/n$.  On the other hand, the
assumption that the eight rises between 3 and 7 kpc are due to caustic
rings  $n = 6+s ... 13 + s$, where $s$ is an integer other than 0 or 1,
yields considerably worse fits.  Up to this point it is unclear whether
$s=0$ or 1 is preferred.  However the two less distinct rises between
7 kpc and $r_\odot$ strongly suggest $s = 1$ since their $r_1$ values
agree at the 2.6\% level with ring radii $a_5^{\rm I}$ and $a_6^{\rm I}$,
but do not agree well with $a_4^{\rm II}$ and $a_5^{\rm II}$. Henceforth
$s = 1$ will be assumed.

The velocity increase due to a caustic ring is given by
\begin{equation}
\Delta v_n = v_{\rm rot} f_n {\Delta I(\zeta_n)
\over \cos\delta_n(0) + \phi_n^\prime(0) \sin\delta_n(0)}\ .
\label{dv}
\end{equation}
The $f_n$, defined in ref. \cite{cr}, are predicted by the self-similar
infall model, but $\Delta I(\zeta_n), \delta_n(0)$ and $\phi_n^\prime(0)$,
defined in ref. \cite{sing}, are not.  Like the $p_n$, the latter
parameters depend in a complicated way on the velocity distribution
of the dark matter at last turnaround.  On the basis of the discussion
in ref. \cite{sing}, the ratio on the RHS of Eq. \ref{dv} is expected
to be of order one, but to vary from one caustic ring to the next.  The
size of these fluctuations is easily a factor two, up or down.  The
sixth column of Table I shows $\Delta v_n$ with the fluctuating ratio
set equal to one, i.e. $\bar{\Delta} v_n \equiv f_n v_{\rm rot}$.

For the reasons just stated, the fact that the observed $\Delta v$
fluctuate by a factor of order 2 from one rise to the next is
consistent with the interpretation that the rises are due to
caustic rings.  However the observed $\Delta v$ (column 3) are
typically a factor 5 larger than the velocity increases caused
by the caustic rings acting alone (column 6).  To account for
the discrepancy I assume that the effect of the caustic rings
is amplified by baryonic matter they have accreted.  First I'll
argue that the gas in the disk has sufficiently high density
and low velocity dispersion for such an explanation to be
plausible.  Second I'll give observational evidence in support
of the explanation.

The equilibrium distribution of gas is:
\begin{equation}
d_{\rm gas} (\vec{r}) = d_{\rm gas}(\vec{r}_0)
\exp[- {3 \over <v^2_{\rm gas}>}(\phi(\vec{r}) - \phi(\vec{r}_0))]\ ,
\label{gas}
\end{equation}
where $d$ is density and $\phi$ gravitational potential.  In the
solar neighborhood, $d_{\rm gas} \simeq
3\cdot 10^{-24} {{\rm gr} \over {\rm cm}^3}$ \cite{BT}, which is
comparable to the density of dark matter inside the tubes of
caustic rings near us.  From the scale height of the gas \cite{BT}
and the assumption that it is in equilibrium with itself and the
other disk components, I estimate
$\langle v_{\rm gas}^2 \rangle ^{1 \over 2} \simeq 8$ km/s.  The
variation in the gravitational potential due to a caustic ring over
the size of the tube is of order $\Delta \phi_{\rm CR} \simeq
2 f v_{\rm rot}^2 {p/a} \simeq (5 {{\rm km} \over {\rm s}})^2$.
Because ${3 \over <v_{\rm gas}^2>} \Delta \phi_{\rm CR}$ is of order
one, the caustic rings have a large effect on the distribution of
gas in the disk.  The accreted gas amplifies and can dominate the
effect of the caustic rings on the rotation curve.  To check
whether this hypothesis is consistent with the shape of the
rises would require detailed modeling, as well as detailed
knowledge on how the rotation curve is measured.  In the
meantime, I found observational evidence in its support.

The accreted gas may reveal the location of caustic rings in
maps of the sky.  Looking tangentially to a ring caustic from
a vantage point in the plane of the ring, one may recognize the
tricusp \cite{sing} shape of the $D_{-4}$ catastrophe.  The IRAS
map of the galactic disk in the direction of galactic coordinates
$(l,b) = (80^\circ, 0^\circ)$ shows a triangular shape which is
strikingly reminiscent of the cross-section of a ring caustic.
The vertices of the triangle are at $(83.5^\circ, 0.3^\circ),
(77.3^\circ, 3.5^\circ)$ and $(77.4^\circ, -2.7^\circ)$ galactic
coordinates.  Images can be obtained from the Skyview Virtual
Observatory (http://skyview.gsfc.nasa.gov/).  The shape is
correctly oriented with respect to the galactic plane and the
galactic center.  Moreover its position is consistent with that
of a rise in the rotation curve, the one between 8.28 and 8.38 kpc
($n=5$ in fit I).  The caustic ring radius implied by the image
is 8.31 kpc, and its dimensions are $p = 134$ pc and $q = 200$ pc,
in the directions parallel and perpendicular to the galactic plane
respectively.

In principle, the feature at $(80^\circ, 0^\circ)$ should be matched
by another in the opposite tangent direction to the nearby ring caustic,
at approximately $(-80^\circ, 0^\circ)$.  Although there is a plausible
feature there, it is much less compelling than the one in the
$(+80^\circ, 0^\circ)$ direction.  There are several reasons why
it may not appear as strongly.  One is that the $(+80^\circ, 0^\circ)$
feature is in the middle of the Sagittarius spiral arm, whose stellar
activity enhances the local gas emissivity, whereas the
$(-80^\circ, 0^\circ)$ feature is not so favorably located.  Another is
that the ring caustic in the $(+80^\circ, 0^\circ)$ direction has
unusually small dimensions.  This may make it more visible by increasing
its contrast with the background.  In the $(-80^\circ,0^\circ)$ direction,
the nearby ring caustic may have larger transverse dimensions.

\section{The big flow} 

Our proximity to a ring means that the associated flows, i.e. those
flows in which the caustic occurs, contribute very importantly to the
local dark matter density.  Using the results of refs. \cite{cr,sing,sty},
and assuming axial symmetry of the caustic ring between us and the
tangent point (approx. 1 kpc away from us), the densities and velocity
vectors on Earth of the associated flows can be derived:
\begin{equation}
d^+ = 1.7~10^{-24}~{{\rm gr} \over {\rm cm}^3}~,~
d^- = 1.5~10^{-25}~{{\rm gr} \over {\rm cm}^3}~,~
\vec{v}^\pm = (470~\hat{\phi} \pm ~100~\hat{r})~
{{\rm km} \over {\rm s}},
\label{lc}
\end{equation}
where $\hat{r}, \hat{\phi}$ and $\hat{z}$ are the local unit vectors
in galactocentric cylindrical coordinates.  $\hat{\phi}$ is in the 
direction of galactic rotation.  The velocities are given in the
(non-rotating) rest frame of the Galaxy.  Because of an ambiguity, 
it is not presently possible to say whether $d^\pm$ are the densities 
of the flows with velocity $\vec{v}^\pm$ or $\vec{v}^\mp$.  Eq. \ref{lc}
has implications for dark matter searches.  Previous estimates of the
local dark matter density, based on isothermal halo profiles, range 
from 5 to 7.5~$10^{-25}~{{\rm gr} \over {\rm cm}^3}$. The present 
analysis implies that a single flow ($d^+$) has three times that 
much local density, i.e. that the total local density is four times 
higher than previously thought.  The large size of $d^+$ is due
to our proximity to a cusp of the nearby caustic.  Assuming axial
symmetry, that cusp is only 55 pc away from us.  The exact size of
$d^+$ is sensitive to our distance to the cusp but, in any case,
$d^+$ is very large.  If we are inside the tube of the fifth caustic,
there are two additional flows on Earth, aside from those given in
Eq. \ref{lc}.  A list of approximate local densities and velocity
vectors for the $n \neq 5$ flows can be found in ref. \cite{bux}.
An updated list is in preparation.

The sharpness of the rises in the rotation curve and of the triangular
feature in the IRAS map implies an upper limit on the velocity dispersion
$\delta v_{\rm DM}$ of the infalling dark matter.  Caustic ring
singularities are spread over a distance of order
$\delta a \simeq {R~\delta v_{\rm DM} \over v}$ where $v$ is the velocity
of the particles in the caustic, $\delta v_{\rm DM}$ is their velocity
dispersion when they first fell in, and $R$ is the turnaround radius
then.  The sharpness of the IRAS feature implies that its edges are
spread over $\delta a \ltwid 20$pc.  Assuming that the feature is
due to the $n=5$ ring caustic, $R \simeq$ 180 kpc and
$v \simeq 480$ km/s.  Therefore $\delta v_{\rm DM} \ltwid 53$ m/s.

There may be evidence for the accretion of baryonic matter onto the
$n \leq 4$ rings as well. Binney and Dehnen studied \cite{bin} the
outer rotation curve of the Milky Way and concluded that its anomalous
behaviour can be explained if most of the tracers of the rotation are
concentrated in a ring of radius $1.6~r_\odot = 13.6$ kpc. This is
very close to the expected radius (13.9 kpc) of the $n=3$ ring.
Recently, the SDSS collaboration detected \cite{hei} an overdensity of
stars which appears to be lying in the galactic plane on an arc of
circle at least $40^\circ$ in length, and of galactocentric radius
approximately 18 kpc.  These stars have properties consistent with
those of spheroid stars but their spatial distribution is not
consistent with a power law spheroid.  The observed feature may
be due to the accretion of stars onto the $n=2$ ring.

The caustic ring model may explain the puzzling persistence of galactic
disk warps \cite{war}.  These may be due to outer caustic rings lying
somewhat outside the galactic plane and attracting visible matter.  Such
disk warps would not damp and would persist on cosmological time scales.

The caustic ring model, and more specifically the prediction Eq.~\ref{lc}
of the locally dominant flow associated with the nearby ring, has
important consequences for axion dark matter searches \cite{adm},
the annual modulation \cite{bux,ann,stiff} and signal anisotropy
\cite{anis,stiff} in WIMP searches, the search for $\gamma$-rays
from dark matter annihilation \cite{gam}, and the search for
gravitational lensing by dark matter caustics \cite{lens}.  The
model allows precise predictions to be made in each of these
approaches to the dark matter problem.


\begin{table}
\caption{Radii at which rises in the Milky Way rotation curve start
($r_1$) and end ($r_2$), the corresponding increases in velocity
$\Delta v$, the caustic ring radii $a_n$ of the self-similar infall
model in the two different fits (I and II) discussed in the text,
and typical velocity increases $\bar{\Delta} v_n$ predicted by the
model, in fit I, without amplification due to baryon accretion.}
\vskip 1cm
\centering
\begin{tabular}{cccccc}
$r_1$&$r_2$&$\Delta v$&$a_n^{\rm I}$&$a_n^{\rm II}$&$\bar{\Delta}v_n$\\
(kpc) & (kpc) & (km/s)     & (kpc)         & (kpc)          &
(km/s)      \\
      &       &            & n = 1 ..14    & n = 1 .. 13    & n = 1 .. 14
\\
\hline
       &       &       &      41.2         &                & 26.5  \\
       &       &       &      20.5         &     37.2       & 10.6  \\
       &       &       &      13.9         &     18.6       &  6.8  \\
       &       &       &      10.5         &     12.5       &  5.0  \\
(8.28) & (8.38)&  (12) &       8.50        &      9.51      &  3.9  \\
(7.30) & (7.42)&   (8) &       7.14        &      7.68      &  3.2  \\
6.24   & 6.84  &   23  &       6.15        &      6.45      &  2.6  \\
5.78   & 6.01  &    9  &       5.41        &      5.56      &  2.3  \\
4.91   & 5.32  &   15  &       4.83        &      4.89      &  2.0  \\
4.18   & 4.43  &    8  &       4.36        &      4.36      &  1.7  \\
3.89   & 4.08  &    8  &       3.98        &      3.94      &  1.5  \\
3.58   & 3.75  &    6  &       3.66        &      3.60      &  1.3  \\
3.38   & 3.49  &   14  &       3.38        &      3.31      &  1.2  \\
3.16   & 3.25  &    8  &       3.15        &      3.05      &  1.1  \\
\hline
\end{tabular}
\label{tbl1}
\end{table}

\end{document}